\documentclass[11pt]{article}
\usepackage{cite}
\usepackage{mathrsfs}
\usepackage{amssymb,amsfonts,amsmath}
\usepackage{enumerate}
\usepackage{indentfirst}
\usepackage{latexsym,bm}
\usepackage[noend]{algpseudocode}
\usepackage{algorithmicx,algorithm}
\usepackage{algorithm}
\usepackage{makeidx}
\usepackage{fancybox}
\usepackage{color}
\usepackage{multicol}
\usepackage[latin1]{inputenc}
\usepackage{graphicx}
\usepackage{pstricks,pst-node,pst-text,pst-3d}
\usepackage{tikz}
\newtheorem{thm}{Theorem}[section]
\newtheorem{cor}[thm]{Corollary}
\newtheorem{lem}[thm]{Lemma}
\newtheorem{op}[thm]{Problem}

\newtheorem{pro}[thm]{Proposition}

\newtheorem{conj}{Conjecture}

\newcommand{\2}{\vspace{0.2cm}}

\begin{document}

\title{\bf Strong Subgraph Connectivity of Digraphs: A Survey}
\author{Yuefang Sun$^{1}$ and Gregory Gutin$^{2}$ \\
$^{1}$ Department of Mathematics,
Shaoxing University\\
Zhejiang 312000, P. R. China, yuefangsun2013@163.com\\
$^{2}$ School of Computer Science and Mathematics\\
Royal Holloway, University of London\\
Egham, Surrey, TW20 0EX, UK, g.gutin@rhul.ac.uk}

\date{}
\maketitle

\begin{abstract}

In this survey we overview known results on the strong subgraph
$k$-connectivity and strong subgraph $k$-arc-connectivity of
digraphs. After an introductory section, the paper is divided
into four sections: basic results, algorithms and complexity, sharp
bounds for strong subgraph $k$-(arc-)connectivity, minimally strong
subgraph $(k, \ell)$-(arc-) connected digraphs. This survey
contains several conjectures and open problems for further study.
\vspace{0.3cm}
\vspace{0.3cm}
\end{abstract}


\section{Introduction}\label{sec:intro}

The generalized $k$-connectivity $\kappa_k(G)$ of a graph $G=(V,E)$
was introduced by Hager \cite{Hager} in 1985 ($2\le k\le |V|$). For
a graph $G=(V,E)$ and a set $S\subseteq V$ of at least two vertices,
an {\em $S$-Steiner tree} or, simply, an {\em $S$-tree}
 is a subgraph
$T$ of $G$ which is a tree with $S\subseteq V(T)$. Two $S$-trees
$T_1$ and $T_2$ are said to be {\em edge-disjoint} if $E(T_1)\cap
E(T_2)=\emptyset$. Two edge-disjoint $S$-trees $T_1$ and $T_2$ are
said to be {\em internally disjoint} if $V(T_1)\cap V(T_2)=S$. The
{\em generalized local connectivity} $\kappa_S(G)$ is the maximum
number of internally disjoint $S$-trees in $G$. For an integer $k$
with $2\leq k\leq n$, the {\em generalized $k$-connectivity} is
defined as
$$\kappa_k(G)=\min\{\kappa_S(G)\mid S\subseteq V(G), |S|=k\}.$$
Observe that $\kappa_2(G)=\kappa(G)$. 
Li, Mao and Sun \cite{Li-Mao-Sun} introduced the
following concept of generalized $k$-edge-connectivity. The {\em
generalized local edge-connectivity} $\lambda_S(G)$ is the maximum
number of edge-disjoint $S$-trees in $G$. For an integer $k$ with
$2\leq k\leq n$, the {\em generalized $k$-edge-connectivity} is
defined as
$$\lambda_k(G)=\min\{\lambda_S(G)\mid S\subseteq V(G), |S|=k\}.$$
Observe that $\lambda_2(G)=\lambda(G)$. Generalized connectivity of
graphs has become an established area in graph theory, see a recent
monograph \cite{Li-Mao5} by Li and Mao on generalized connectivity
of undirected graphs.

To extend generalized $k$-connectivity to directed graphs, Sun,
Gutin, Yeo and Zhang \cite{Sun-Gutin-Yeo-Zhang} observed that in the
definition of $\kappa_S(G)$, one can replace ``an $S$-tree'' by ``a
connected subgraph of $G$ containing $S$.'' Therefore,  Sun et al.
\cite{Sun-Gutin-Yeo-Zhang} defined {\em strong subgraph
$k$-connectivity} by replacing ``connected'' with ``strongly
connected'' (or, simply, ``strong'') as follows. Let $D=(V,A)$ be a
digraph of order $n$, $S$ a subset of $V$ of size $k$ and $2\le
k\leq n$. A subdigraph $H$ of $D$ is called an {\em $S$-strong
subgraph} if $H$ is strong and $S\subseteq V(H)$. Two $S$-strong
subgraphs $D_1$ and $D_2$ are said to be {\em arc-disjoint} if
$A(D_1)\cap A(D_2)=\emptyset$. Two arc-disjoint $S$-strong subgraphs
$D_1$ and $D_2$ are said to be {\em internally disjoint} if
$V(D_1)\cap V(D_2)=S$. Let $\kappa_S(D)$ be the maximum number of
internally disjoint $S$-strong subgraphs in $D$. The {\em strong
subgraph $k$-connectivity} of $D$ is defined as
$$\kappa_k(D)=\min\{\kappa_S(D)\mid S\subseteq V, |S|=k\}.$$
By definition, $\kappa_k(D)=0$ if $D$ is not strong.

As a natural counterpart of the strong subgraph $k$-connectivity,
Sun and Gutin \cite{Sun-Gutin2} introduced the concept of strong
subgraph $k$-arc-connectivity. Let $D=(V(D),A(D))$ be a digraph of
order $n$, $S\subseteq V$ a $k$-subset of $V(D)$ and $2\le k\leq n$.
Let $\lambda_S(D)$ be the maximum number of arc-disjoint $S$-strong
digraphs in $D$. The {\em strong subgraph $k$-arc-connectivity} of
$D$ is defined as
$$\lambda_k(D)=\min\{\lambda_S(D)\mid S\subseteq V(D), |S|=k\}.$$
By definition, $\lambda_k(D)=0$ if $D$ is not strong.

The strong subgraph $k$-(arc-)connectivity is not only a natural
extension of the concept of generalized $k$-(edge-)connectivity, but
also relates to important problems in graph theory. For $k=2$,
$\kappa_2(\overleftrightarrow{G})=\kappa(G)$
\cite{Sun-Gutin-Yeo-Zhang} and
$\lambda_2(\overleftrightarrow{G})=\lambda(G)$ \cite{Sun-Gutin2}.
Hence, $\kappa_k(D)$ and $\lambda_k(D)$ could be seen as
generalizations of connectivity and edge-connectivity of undirected
graphs, respectively. For $k=n$, $\kappa_n(D)=\lambda_n(D)$ is the
maximum number of arc-disjoint spanning strong subgraphs of $D$.
Moreover, since $\kappa_S(G)$ and $\lambda_S(G)$ are the
number of internally disjoint and arc-disjoint strong subgraphs
containing a given set $S$, respectively, these parameters
are relevant to the subdigraph packing problem, see
\cite{Bang-Jensen-Huang, Bang-Jensen-Kriesell, Bang-Jensen-Yeo,
Bang-Jensen-Yeo2, Cheriyan-Salavatipour}.

Some basic results will be introduced in Section \ref{sec:basic}. In
Section \ref{sec:complexity}, we will sum up the results on
algorithms and computational complexity for $\kappa_S(D)$,
$\kappa_k(D)$, $\lambda_S(D)$ and $\lambda_k(D)$. We will collect
many upper and lower bounds for the parameters $\kappa_k(D)$ and
$\lambda_k(D)$ in Section \ref{sec:bounds}. Finally, in Section
\ref{sec:mini}, results on minimally strong subgraph $(k,
\ell)$-(arc-)connected digraphs will be surveyed.

\paragraph{Additional Terminology and Notation.} For a digraph $D$, its {\em reverse} $D^{\rm rev}$ is a digraph with same vertex set and such that
$xy\in A(D^{\rm rev})$ if and only if $yx\in A(D)$. A digraph $D$ is
{\em symmetric} if $D^{\rm rev}=D$. In other words, a symmetric
digraph $D$ can be obtained from its underlying undirected graph $G$
by replacing each edge of $G$ with the corresponding arcs of both
directions, that is, $D=\overleftrightarrow{G}.$ A 2-cycle $xyx$ of
a strong digraph $D$ is called a {\em bridge} if $D-\{xy,yx\}$ is
disconnected. Thus, a bridge corresponds to a bridge in the
underlying undirected graph of $D$. An {\em orientation} of a
digraph $D$ is a digraph obtained from $D$ by deleting an arc in
each 2-cycle of $D$. A digraph $D$ is {\em semicomplete} if for
every distinct $x,y\in V(D)$ at least one of the arcs $xy,yx$ in in
$D$. 
We refer the readers to
\cite{Bang-Jensen-Gutin, Bang-Jensen-Gutin2, Bondy} for graph
theoretical notation and terminology not given here.


\section{Basic Results}\label{sec:basic}

The following proposition can be easily verified using definitions
of $\lambda_{k}(D)$ and $\kappa_k(D)$.

\begin{pro}\label{pro00}\cite{Sun-Gutin-Yeo-Zhang, Sun-Gutin2}
Let $D$ be a digraph of order $n$, and let $k\ge 2$ be an integer.
Then
\begin{equation}\label{pro01}
\lambda_{k+1}(D)\leq \lambda_{k}(D) \mbox{ for every } k\le n-1
\end{equation}
\begin{equation}\label{pro02}
\kappa_k(D')\leq \kappa_k(D), \lambda_k(D')\leq \lambda_k(D) \mbox{
where $D'$ is a spanning subdigraph of $D$}
\end{equation}
\begin{equation}\label{pro03}
\kappa_k(D)\leq \lambda_k(D) \leq \min\{\delta^+(D), \delta^-(D)\}
\end{equation}
\end{pro}

By Tillson's decomposition theorem \cite{Tillson}, we can
determine the exact values for $\kappa_k(\overleftrightarrow{K}_n)$
and $\lambda_k(\overleftrightarrow{K}_n)$.

\begin{pro}\label{pro04}\cite{Sun-Gutin-Yeo-Zhang} For $2\leq k\leq n$, we have

\[
\kappa_k(\overleftrightarrow{K}_n)=\left\{
   \begin{array}{ll}
      {n-1}, & \mbox{if $k\not\in \{4,6\}$;}\\
     {n- 2}, &\mbox{otherwise.}
   \end{array}
   \right.
\]
\end{pro}

\begin{pro}\label{pro05}\cite{Sun-Gutin2} For $2\leq k\leq n$, we have

\[ \lambda_k(\overleftrightarrow{K}_n)=\left\{
   \begin{array}{ll}
      {n-1}, & \mbox{if $k\not\in \{4,6\}$,~or,~$k\in \{4,6\}$~and~$k<n$;}\\
     {n- 2}, &\mbox{if $k=n\in \{4,6\}$.}
   \end{array}
   \right.
\]
\end{pro}

\begin{pro}\label{pro07}\cite{Sun-Gutin2}
For every fixed $k\ge 2$, a digraph $D$ is strong if and only if $\lambda_k(D)\ge 1.$
\end{pro}


\section{Algorithms and Complexity}\label{sec:complexity}

\subsection{Results for $\kappa_S(D)$ and $\kappa_k(D)$}

For a fixed $k\ge 2$, it is easy to decide whether $\kappa_k(D)\ge 1$ for a digraph $D$:
it holds if and only if $D$ is strong. Unfortunately, deciding
whether $\kappa_S(D)\ge 2$ is already NP-complete for $S \subseteq
V(D)$ with $|S|=k$, where $k\ge 2$ is a fixed integer.

The well-known {\sc Directed $q$-Linkage} problem was proved to be
NP-complete even for the case that $q=2$
\cite{Fortune-Hopcroft-Wyl}. The problem is formulated as follows:
for a fixed integer $q\ge 2$, given a digraph $D$ and a (terminal)
sequence $((s_1,t_1),\dots ,(s_q,t_q))$ of distinct vertices of $D,$
decide whether $D$ has $q$ vertex-disjoint paths $P_1,\dots ,P_q$,
where $P_i$ starts at $s_i$ and ends at $t_i$ for all $i\in [q].$

By using the reduction from the {\sc Directed $q$-Linkage} problem,
we can prove the following intractability result.

\begin{thm}\label{thm01}\cite{Sun-Gutin-Yeo-Zhang} Let $k\ge 2$ and $\ell\ge 2$ be fixed integers.
Let $D$ be a digraph and $S \subseteq V(D)$ with $|S|=k$. The
problem of deciding whether $\kappa_S(D)\ge \ell$ is NP-complete.
\end{thm}

In the above theorem, Sun et al. obtained the complexity result of
the parameter $\kappa_S(D)$ for an arbitrary digraph $D$. For
$\kappa_k(D)$, they made the following conjecture.

\begin{conj}\label{conj1}\cite{Sun-Gutin-Yeo-Zhang}
It is NP-complete to decide for fixed integers $k\ge 2$ and $\ell\ge
2$ and a given digraph $D$ whether $\kappa_k(D)\ge \ell$.
\end{conj}

Recently, Chudnovsky, Scott and Seymour \cite{Chud-Scott-Seymour}
proved the following powerful result.

\begin{thm}\label{thm:CSS}\cite{Chud-Scott-Seymour}
Let $q$ and $c$ be fixed positive integers. Then the  {\sc Directed
$q$-Linkage} problem on a digraph $D$ whose vertex set can be
partitioned into $c$ sets each inducing a semicomplete digraph and a
terminal sequence $((s_1,t_1),\dots ,(s_q,t_q))$ of  distinct
vertices of $D$, can be solved in polynomial time.
\end{thm}

The following nontrivial lemma can be deduced from Theorem
\ref{thm:CSS}.

\begin{lem}\label{lem01}\cite{Sun-Gutin-Yeo-Zhang}
Let $k$ and $\ell$ be fixed positive integers. Let $D$ be a digraph
and let $X_1,X_2,\ldots,X_{\ell}$ be $\ell$ vertex disjoint subsets
of $V(D)$, such that $|X_i| \leq k$ for all $i\in [\ell]$. Let $X =
\cup_{i=1}^{\ell} X_i$ and assume that every vertex in $V(D)
\setminus X$ is adjacent to every other vertex in $D$. Then we can
in polynomial time decide if there exists vertex disjoint subsets
$Z_1,Z_2,\ldots,Z_{\ell}$ of $V(D)$, such that $X_i \subseteq Z_i$
and $D[Z_i]$ is strongly connected for each $i\in [\ell]$.
\end{lem}

Using Lemma \ref{lem01}, Sun, Gutin, Yeo and Zhang proved the following
result for semicomplete digraphs.
\begin{thm}\label{thm02}\cite{Sun-Gutin-Yeo-Zhang}
For any fixed integers $k, \ell \ge 2$, we can decide whether
$\kappa_k(D)\ge \ell$ for a semicomplete digraph $D$ in polynomial
time.
\end{thm}

Now we turn our attention to symmetric graphs. We start with the
following structural result.

\begin{thm}\label{thm03}\cite{Sun-Gutin-Yeo-Zhang}
For every undirected graph $G$ we have
$\kappa_2(\overleftrightarrow{G})=\kappa(G)$.
\end{thm}

Theorem~\ref{thm03} immediatly implies the following positive
result, which follows from the fact that $\kappa(G)$ can be computed
in polynomial time.

\begin{cor}\label{cor01}\cite{Sun-Gutin-Yeo-Zhang}
For a graph $G$, $\kappa_2(\overleftrightarrow{G})$ can be computed
in polynomial time.
\end{cor}

Theorem~\ref{thm03} states that
$\kappa_k(\overleftrightarrow{G})=\kappa_k(G)$ when $k=2$. However
when $k \geq 3$, then $\kappa_k(\overleftrightarrow{G})$ is not
always equal to $\kappa_k(G)$, as can be seen from
$\kappa_3(\overleftrightarrow{K_3}) = 2 \not= 1 = \kappa_3(K_3)$.
Chen, Li, Liu and Mao \cite{Chen-Li-Liu-Mao} introduced the
following problem, which they proved to be NP-complete.

\2

{\sc CLLM Problem:} Given a tripartite graph $G=(V, E)$ with a
3-partition $(\overline{U}, \overline{V}, \overline{W})$ such that
$|\overline{U}|=|\overline{V}|=|\overline{W}|=q$, decide whether
there is a partition of $V$ into $q$ disjoint 3-sets $V_1, \dots,
V_q$ such that for every $V_i= \{v_{i_1}, v_{i_2}, v_{i_3}\}$
$v_{i_1} \in \overline{U}, v_{i_2} \in \overline{V}, v_{i_3} \in
\overline{W}$ and $G[V_i]$ is connected.

\begin{lem}\label{lem:CLLM}\cite{Chen-Li-Liu-Mao} The CLLM Problem is NP-complete.
\end{lem}

Now restricted to symmetric digraphs $D$, for any fixed integer
$k\geq 3$, by Lemma \ref{lem:CLLM}, the problem of deciding whether
$\kappa_S(D)\geq \ell~(\ell \geq 1)$ is NP-complete for $S\subseteq
V(D)$ with $|S|=k$.

\begin{thm}\label{thm04}\cite{Sun-Gutin-Yeo-Zhang}
For any fixed integer $k\geq 3$, given a symmetric digraph $D$, a
$k$-subset $S$ of $V(D)$ and an integer $\ell~(\ell \geq 1)$,
deciding whether $\kappa_S(D)\geq \ell$, is NP-complete.
\end{thm}

The last theorem assumes that $k$ is fixed but $\ell$ is a part of
input. When both $k$ and $\ell$ are fixed, the problem of deciding
whether $\kappa_S(D) \geq \ell$ for a symmetric digraph $D$, is
polynomial-time solvable. We will start with the following technical
lemma.

\begin{lem}\label{lem02}\cite{Sun-Gutin-Yeo-Zhang}
Let $k,\ell \geq 2$ be fixed. Let $G$ be a graph and let $S
\subseteq V(G)$ be an independent set in $G$ with $|S|=k$. For $i\in
[\ell]$, let $D_i$ be any set of arcs with both end-vertices in $S$.
Let a forest $F_i$ in $G$ be called $(S,D_i)$-{\em acceptable} if
the digraph $\overleftrightarrow{F_i}+D_i$ is strong and contains
$S$. In polynomial time, we can decide whether there exists
edge-disjoint forests $F_1,F_2,\ldots,F_{\ell}$ such that $F_i$ is
$(S,D_i)$-acceptable for all $i\in [\ell]$ and $V(F_i) \cap V(F_j)
\subseteq S$ for all $1 \leq i < j \leq \ell$.
\end{lem}

Now we can prove the following result by Lemma \ref{lem02}:

\begin{thm}\label{thm05}\cite{Sun-Gutin-Yeo-Zhang}
Let $k, \ell \geq 2$ be fixed. For any symmetric digraph $D$ and $S
\subseteq V(D)$ with $|S|=k$ we can in polynomial time decide
whether $\kappa_S(D) \geq \ell$.
\end{thm}

The {\sc Directed $q$-Linkage} problem is polynomial-time solvable
for planar digraphs \cite{Schr} and digraphs of bounded directed
treewidth \cite{JRST}. However, it seems that we cannot use the
approach in proving Theorem \ref{thm02} directly as the structure of
minimum-size strong subgraphs in these two classes of digraphs is
more complicated than in semicomplete digraphs. Certainly, we cannot
exclude the possibility that computing strong subgraph
$k$-connectivity in planar digraphs and/or in digraphs of bounded
directed treewidth is NP-complete.

\begin{op}\label{op1}\cite{Sun-Gutin-Yeo-Zhang}
What is the complexity of deciding whether $\kappa_k(D)\ge \ell$ for
fixed integers $k\ge 2$, and $\ell\ge 2$ and a given planar digraph $D$?
\end{op}

\begin{op}\label{op5}\cite{Sun-Gutin-Yeo-Zhang}
What is the complexity of deciding whether $\kappa_k(D)\ge \ell$ for
fixed integers $k\ge 2$, and $\ell\ge 2$ and a digraph $D$ of bounded
directed treewidth?
\end{op}

It would be interesting to identify large classes of digraphs for which the $\kappa_k(D)\ge \ell$ problem can be decided  in polynomial time.

\subsection{Results for $\lambda_S(D)$ and $\lambda_k(D)$}

Yeo proved that it is an NP-complete problem to decide whether a
2-regular digraph has two arc-disjoint hamiltonian cycles (see,
e.g., Theorem 6.6 in \cite{Bang-Jensen-Yeo}). (A digraph is 2-regular if the out-degree and in-degree
of every vertex equals 2.) Thus, the problem of
deciding whether $\lambda_n(D)\ge 2$ is NP-complete, where $n$ is
the order of $D$. Sun and Gutin \cite{Sun-Gutin2} extended this
result in Theorem \ref{thm07}.

Let $D$ be a digraph and let
$s_1,s_2,\ldots{},s_q,t_1,t_2,\ldots{},t_q$ be a collection of not
necessarily distinct vertices of $D$.
 A {\em weak $q$-linkage} from $(s_1,s_2,\ldots{},s_q)$ to $(t_1,t_2,\ldots{},t_q)$ is a collection of $q$ arc-disjoint paths
 $P_1,\ldots{},P_q$ such that $P_i$ is
an $(s_i,t_i)$-path for each $i\in [q]$. A digraph $D=(V,A)$ is {\em
weakly $q$-linked} if it contains a weak $q$-linkage
 from $(s_1,s_2,\ldots{},s_q)$ to $(t_1,t_2,\ldots{},t_q)$ for every choice of (not necessarily
distinct) vertices $s_1,\ldots{},s_q,t_1,\ldots{},t_q$. The {\sc
Directed Weak $q$-Linkage} problem is the following. Given a digraph
$D=(V,A)$ and  distinct vertices $x_1,x_2,\ldots{},x_q,
y_1,y_2,\ldots{},y_q$; decide whether $D$ contains $q$ arc-disjoint
paths $P_1,\ldots{},P_q$ such that $P_i$ is an $(x_i,y_i)$-path. The
problem is well-known to be NP-complete already for $q=2$
\cite{Fortune-Hopcroft-Wyl}. By using the reduction from the {\sc
Directed Weak $q$-Linkage} problem, we can prove the following
intractability result.

\begin{thm}\label{thm07}\cite{Sun-Gutin2}
Let $k\ge 2$ and $\ell\ge 2$ be fixed integers. Let $D$ be a digraph
and $S \subseteq V(D)$ with $|S|=k$. The problem of deciding whether
$\lambda_S(D)\ge \ell$ is NP-complete.
\end{thm}

Bang-Jensen and Yeo \cite{Bang-Jensen-Yeo} conjectured the
following:

\begin{conj}\label{conj2}
For every $\lambda\ge 2$ there is a finite set ${\cal S}_{\lambda}$
of digraphs such that $\lambda$-arc-strong semicomplete digraph $D$
contains $\lambda$ arc-disjoint spanning strong subgraphs unless
$D\in {\cal S}_{\lambda}$.
\end{conj}

Bang-Jensen and Yeo \cite{Bang-Jensen-Yeo} proved the conjecture for
$\lambda=2$ by showing that $|{\cal S}_2|=1$ and describing the
unique digraph $S_4$ of  ${\cal S}_2$ of order 4. This result and
Theorem \ref{thm13} imply the following:

\begin{thm}\label{thm08}\cite{Sun-Gutin2}
For a semicomplete digraph $D$,  of order $n$ and an integer $k$
such that $2\le k\le n$, $\lambda_k(D)\ge 2$ if and only if $D$ is
2-arc-strong and $D\not\cong S_4$.
\end{thm}

Now we turn our attention to symmetric graphs. We start from
characterizing symmetric digraphs $D$ with $\lambda_k(D)\ge 2$, an
analog of Theorem \ref{thm08}. To prove it we need the following
result of Boesch and Tindell \cite{BT} translated from the language
of mixed graphs to that of digraphs.

\begin{thm}\label{thm:BT}
A strong digraph $D$ has a strong orientation if and only if $D$ has
no bridge.
\end{thm}

Here is the characterization by Sun and Gutin.

\begin{thm}\label{thm09}\cite{Sun-Gutin2}
For a strong symmetric digraph $D$ of order $n$ and an integer $k$
such that $2\le k\le n$, $\lambda_k(D)\ge 2$ if and only if $D$ has
no bridge.
\end{thm}

Theorems \ref{thm08} and \ref{thm09} imply the following complexity
result, which we believe to be extendable from $\ell=2$ to any
natural $\ell\ge 2$.

\begin{cor}\label{cor02}\cite{Sun-Gutin2}
The problem of deciding whether $\lambda_k(D)\ge 2$ is
polynomial-time solvable if $D$ is either semicomplete or symmetric
digraph of order $n$ and $2\le k\le n. $
\end{cor}

Sun and Gutin gave a lower bound on $\lambda_k(D)$ for symmetric
digraphs $D$.

\begin{thm}\label{thm10}\cite{Sun-Gutin2}
For every graph $G$, we have
$$\lambda_k(\overleftrightarrow{G})\geq \lambda_k(G).$$ Moreover, this
bound is sharp. In particular, we have
$\lambda_2(\overleftrightarrow{G})=\lambda_2(G)$.
\end{thm}

Theorem~\ref{thm10} immediately implies the next result, which
follows from the fact that $\lambda(G)$ can be computed in
polynomial time.

\begin{cor}\label{cor03}\cite{Sun-Gutin2}
For a symmetric digraph $D$, $\lambda_2(D)$ can be computed in
polynomial time.
\end{cor}

Corollaries \ref{cor02} and \ref{cor03} shed some light on the
complexity of deciding, for fixed $k,\ell\ge 2$, whether
$\lambda_k(D)\ge \ell$ for semicomplete and symmetric digraphs $D$.
However, it is unclear what is the complexity above for every fixed
$k,\ell\ge 2$. If Conjecture \ref{conj2} is correct, then the
$\lambda_k(D)\ge \ell$ problem can be solved in polynomial time for
semicomplete digraphs. However, Conjecture \ref{conj2} seems to be
very difficult. It was proved in \cite{Sun-Gutin-Yeo-Zhang} that for
fixed $k, \ell\ge 2$ the problem of deciding whether $\kappa_k(D)\ge
\ell$ is polynomial-time solvable for both semicomplete and
symmetric digraphs, but it appears that the approaches to prove the
two results cannot be used for $\lambda_k(D)$. Some well-known
results such as the fact that the hamiltonicity problem is
NP-complete for undirected 3-regular graphs, indicate that  the
$\lambda_k(D)\ge \ell$ problem for symmetric digraphs may be
NP-complete, too.

\begin{op}\label{op2}\cite{Sun-Gutin2}
What is the complexity of deciding whether $\lambda_k(D)\ge \ell$
for fixed integers $k\ge 2$ and $\ell\ge 2$, and a semicomplete
digraph $D$?
\end{op}

\begin{op}\label{op6}\cite{Sun-Gutin2}
What is the complexity of deciding whether $\lambda_k(D)\ge \ell$
for fixed integers $k\ge 2$ and $\ell\ge 2$, and a symmetric
digraph $D$?
\end{op}

It would be interesting to identify large classes of digraphs for which the $\lambda_k(D)\ge \ell$ problem can be decided  in polynomial time.


\section{Bounds for Strong Subgraph $k$-(Arc-)Connectivity}\label{sec:bounds}

\subsection{Results for $\kappa_k(D)$}

By Propositions \ref{pro00} and \ref{pro04}, Sun, Gutin, Yeo and
Zhang obtained a sharp lower bound and a sharp upper bound for
$\kappa_k(D)$, where $2\leq k\leq n$.

\begin{thm}\label{thm06}\cite{Sun-Gutin-Yeo-Zhang}
Let $2\leq k\leq n$. For a strong digraph $D$ of order $n$, we have
$$1\leq \kappa_k(D)\leq n-1.$$ Moreover, both bounds are sharp, and
the upper bound holds if and only if $D\cong
\overleftrightarrow{K}_n$, $2\leq k\leq n$ and $k\not\in \{4,6\}$.
\end{thm}

Sun and Gutin gave the following sharp upper bound for $\kappa_k(D)$
which improves (3) of Proposition \ref{pro00}.

\begin{thm}\label{thm11}\cite{Sun-Gutin}
For $k\in \{2,\dots ,n\}$ and $n\ge \kappa(D)+k,$ we have
$$\kappa_k(D)\leq \kappa(D).$$ Moreover, the bound is sharp.
\end{thm}

\subsection{Results for $\lambda_k(D)$}

By Propositions \ref{pro00} and \ref{pro04}, Sun and Gutin obtained
a sharp lower bound and a sharp upper bound for $\lambda_k(D)$, where
$2\leq k\leq n$.

\begin{thm}\label{thm12}\cite{Sun-Gutin2}
Let $2\leq k\leq n$. For a strong digraph $D$ of order $n$, we have
$$1\leq \lambda_k(D)\leq n-1.$$ Moreover, both bounds are sharp, and
the upper bound holds if and only if $D\cong
\overleftrightarrow{K}_n$, where $k\not\in \{4,6\}$,~or,~$k\in
\{4,6\}$~and~$k<n$.
\end{thm}

They also gave the following sharp upper bound for $\lambda_k(D)$
which improves (3) of Proposition \ref{pro00}.

\begin{thm}\label{thm13}\cite{Sun-Gutin2}
For $2\leq k\leq n$, we have
$$\lambda_k(D)\leq \lambda(D).$$ Moreover, the bound is sharp.
\end{thm}

Shiloach \cite{shiloachIPL8} proved the following:

\begin{thm}\cite{shiloachIPL8}\label{Shiloach}
A digraph $D$ is weakly $k$-linked if and only if
$D$ is $k$-arc-strong.
\end{thm}

Using Shiloach's Theorem, Sun and Gutin \cite{Sun-Gutin2} proved the
following lower bound for $\lambda_k(D)$. Such a bound does not hold
for $\kappa_k(D)$ since it was shown in \cite{Sun-Gutin-Yeo-Zhang}
using Thomassen's result in \cite{Thom} that for every $\ell$ there
are digraphs $D$ with $\kappa(D)=\ell$ and $\kappa_2(D)=1$.

\begin{pro}\label{pro06}\cite{Sun-Gutin2}
Let $k\le \ell=\lambda(D)$. We have $\lambda_k(D)\ge \lfloor
\ell/k\rfloor $.
\end{pro}

For a digraph $D=(V(D), A(D))$, the {\em complement digraph},
denoted by $D^c$, is a digraph with vertex set $V(D^c)=V(D)$ such
that $xy\in A(D^c)$ if and only if $xy\not\in A(D)$.

Given a graph parameter $f(G)$, the Nordhaus-Gaddum Problem is to
determine sharp bounds for (1) $f(G) + f(G^c)$ and (2) $f(G)f(G^c)$,
and characterize the extremal graphs. The Nordhaus-Gaddum type
relations have received wide attention; see a recent survey paper
\cite{Aouchiche-Hansen} by Aouchiche and Hansen. By using
Proposition \ref{pro07}, the following Theorem \ref{thm14}
concerning such type of a problem for the parameter $\lambda_k$ can
be obtained.

\begin{thm}\label{thm14}\cite{Sun-Gutin2}
For a digraph $D$ with order $n$, the following assertions holds:\\
$(i)$~$0\leq \lambda_k(D)+\lambda_k(D^c)\leq n-1$. Moreover, both bounds are sharp. In particular, the lower bound holds if and only if $\lambda(D)=\lambda(D^c)=0$.\\
$(ii)$~$0\leq \lambda_k(D){\lambda_k(D^c)}\leq (\frac{n-1}{2})^2$.
Moreover, both bounds are sharp. In particular, the lower bound
holds if and only if $\lambda(D)=0$ or $\lambda(D^c)=0$.
\end{thm}


We now discuss Cartesian products of digraphs. The {\em Cartesian
product} $G\Box H$ of two digraphs $G$ and $H$ is a digraph with
vertex set
$$V(G\Box H)=V(G)\times V(H)=\{(x, x')\mid x\in V(G), x'\in V(H)\}$$
and arc set $$A(G\Box H)=\{(x,x')(y,y')\mid xy\in A(G),
x'=y',~or~x=y, x'y'\in A(H)\}.$$ By definition, we know the
Cartesian product is associative and commutative, and $G\Box H$ is
strongly connected if and only if both $G$ and $H$ are strongly
connected \cite{Hammack}.

\begin{thm}\label{thm15}\cite{Sun-Gutin2}
Let $G$ and $H$ be two digraphs. We have $$\lambda_2(G\Box H)\geq
\lambda_2(G)+ \lambda_2(H).$$ Moreover, the bound is sharp.
\end{thm}

\begin{figure}[htbp]
{\tiny
\begin{center}
\renewcommand\arraystretch{3.5}
\begin{tabular}{|p{1.5cm}|p{1.5cm}|p{1.5cm}|p{1.5cm}|p{1.5cm}|}
\hline & $\overrightarrow{C}_m$ & $\overleftrightarrow{C}_m$ &
$\overleftrightarrow{T}_m$ & $\overleftrightarrow{K}_m$
\\\hline

$\overrightarrow{C}_n$ & $2$ & $3$ & $2$ & $m$
\\\hline

$\overleftrightarrow{C}_n$ & $3$ & $4$ & $3$ & $m+1$
\\\hline

$\overleftrightarrow{T}_n$ & $2$ & $3$ & $2$ & $m$
\\\hline

$\overleftrightarrow{K}_n$ & $n$ & $n+1$ & $n$ & $n+m-2$
\\\hline

\end{tabular}
\vspace*{40pt}

\centerline{\normalsize Table $1$. Precise values for the strong
subgraph 2-arc-connectivity of some special cases.}
\end{center}}
\end{figure}

By Proposition \ref{pro00} and Theorem \ref{thm15}, we can obtain
precise values for the strong subgraph 2-arc-connectivity of the
Cartesian product of some special digraphs, as shown in the Table.
Note that $\overleftrightarrow{T}_m$ is the symmetric digraph whose
underlying undirected graph is a tree of order $m$.


\section{Minimally Strong Subgraph $(k,\ell)$-(Arc-)Connected
Digraphs}\label{sec:mini}

\subsection{Results for Minimally Strong Subgraph $(k,\ell)$-Connected
Digraphs}

A digraph $D=(V(D), A(D))$ is called {\em minimally strong subgraph
$(k,\ell)$-connected} if $\kappa_k(D)\geq \ell$ but for any arc
$e\in A(D)$, $\kappa_k(D-e)\leq \ell-1$ \cite{Sun-Gutin}. By the
definition of $\kappa_k(D)$ and Theorem \ref{thm06}, we know $2\leq
k\leq n, 1\leq \ell \leq n-1$. Let $\mathfrak{F}(n,k,\ell)$ be the
set of all minimally strong subgraph $(k,\ell)$-connected digraphs
with order $n$. We define
$$F(n,k,\ell)=\max\{|A(D)| \mid D\in \mathfrak{F}(n,k,\ell)\}$$ and
$$f(n,k,\ell)=\min\{|A(D)| \mid D\in \mathfrak{F}(n,k,\ell)\}.$$ We
further define $$Ex(n,k,\ell)=\{D\mid D\in \mathfrak{F}(n,k,\ell),
|A(D)|=F(n,k,\ell)\}$$ and $$ex(n,k,\ell)=\{D\mid D\in
\mathfrak{F}(n,k,\ell), |A(D)|=f(n,k,\ell)\}.$$

By the definition of a minimally strong subgraph
$(k,\ell)$-connected digraph, we can get the following observation.

\begin{pro}\label{pro08}\cite{Sun-Gutin}
A digraph $D$ is minimally strong subgraph $(k,\ell)$-connected if
and only if $\kappa_k(D)= \ell$ and $\kappa_k(D-e)= \ell-1$ for any
arc $e\in A(D)$.
\end{pro}

A digraph $D$ is {\em minimally strong} if $D$ is strong but $D-e$
is not for every arc $e$ of $D$.

\begin{pro}\label{pro09}\cite{Sun-Gutin}
The following assertions hold:\\
$(i)$~A digraph $D$ is minimally strong subgraph $(k,1)$-connected
if and only if $D$ is minimally strong digraph;\\
$(ii)$~For $k\neq 4,6$, a digraph $D$ is minimally strong subgraph
$(k,n-1)$-connected if and only if $D\cong
\overleftrightarrow{K}_n$.
\end{pro}

The following result characterizes minimally strong subgraph
$(2,n-2)$-connected digraphs.

\begin{thm}\label{thm16}\cite{Sun-Gutin}
A digraph $D$ is minimally strong subgraph $(2,n-2)$-connected if
and only if $D$ is a digraph obtained from the complete digraph
$\overleftrightarrow{K}_n$ by deleting an arc set M such that
$\overleftrightarrow{K}_n[M]$ is a 3-cycle or a union of $\lfloor
n/2\rfloor$ vertex-disjoint 2-cycles. In particular, we have
$f(n,2,n-2)=n(n-1)-2\lfloor n/2\rfloor$, $F(n,2,n-2)=n(n-1)-3$.
\end{thm}

Note that Theorem \ref{thm16} implies that
$Ex(n,2,n-2)=\{\overleftrightarrow{K_n}-M\}$ where $M$ is an arc set
such that $\overleftrightarrow{K}_n[M]$ is a directed 3-cycle, and
$ex(n,2,n-1)=\{\overleftrightarrow{K_n}-M\}$ where $M$ is an arc set
such that $\overleftrightarrow{K}_n[M]$ is a union of $\lfloor
n/2\rfloor$ vertex-disjoint directed 2-cycles.

The following result concerns a sharp lower bound for the parameter
$f(n,k,\ell)$.

\begin{thm}\label{thm17}\cite{Sun-Gutin} For $2\leq k\leq n$, we have
$$f(n,k,\ell)\geq n\ell.$$
Moreover, the following assertions hold:\\
$(i)~$ If $\ell=1$, then $f(n,k,\ell)=n$; $(ii)~$ If $2\leq \ell\leq
n-1$, then $f(n,n,\ell)=n\ell$ for $k=n\not\in \{4,6\}$; (iii) If
$n$ is even and $\ell = n-2$, then $f(n,2,\ell)=n\ell.$
\end{thm}

To prove two upper bounds on the number of arcs in a minimally
strong subgraph $(k,\ell)$-connected digraph, Sun and Gutin used the
following result, see e.g. \cite{Bang-Jensen-Gutin}.

\begin{thm}\label{2n-2-thm}
Every strong digraph $D$ on $n$ vertices has a strong spanning
subgraph $H$ with at most $2n-2$ arcs and equality holds only if $H$
is a symmetric digraph whose underlying undirected graph is a tree.
\end{thm}

\begin{pro}\label{pro10}\cite{Sun-Gutin}
We have $(i)$~$F(n,n,\ell)\le 2\ell(n-1)$; $(ii)$~For every $k$
$(2\le k\le n)$, $F(n,k,1)=2(n-1)$ and $Ex(n,k,1)$ consists of
symmetric digraphs whose underlying undirected graphs are trees.
\end{pro}

The minimally strong subgraph $(2,n-2)$-connected digraphs was
characterized in Theorem \ref{thm16}. As a simple consequence of the
characterization, we can determine the values of $f(n,2,n-2)$ and
$F(n,2,n-2)$. It would be interesting to determine $f(n,k,n-2)$ and
$F(n,k,n-2)$ for every value of $k\ge 3$ since obtaining
characterizations of all $(k,n-2)$-connected digraphs for $k\ge 3$
seems a very difficult problem.

\begin{op}\label{op3}\cite{Sun-Gutin}
Determine $f(n,k,n-2)$ and $F(n,k,n-2)$ for every value of $k\ge 3$.
\end{op}

It would also be interesting to find a sharp upper bound for
$F(n,k,\ell)$ for all $k\ge 2$ and $\ell\ge 2$.

\begin{op}\label{op4}\cite{Sun-Gutin}
Find a sharp upper bound for $F(n,k,\ell)$ for all $k\ge 2$ and
$\ell\ge 2$.
\end{op}

\subsection{Results for Minimally Strong Subgraph $(k,\ell)$-Arc-Connected
Digraphs}

A digraph $D=(V(D), A(D))$ is called {\em minimally strong subgraph
$(k,\ell)$-arc-connected} if $\lambda_k(D)\geq \ell$ but for any arc
$e\in A(D)$, $\lambda_k(D-e)\leq \ell-1$. By the definition of
$\lambda_k(D)$ and Theorem \ref{thm12}, we know $2\leq k\leq n,
1\leq \ell \leq n-1$. Let $\mathfrak{G}(n,k,\ell)$ be the set of all
minimally strong subgraph $(k,\ell)$-arc-connected digraphs with
order $n$. We define
$$G(n,k,\ell)=\max\{|A(D)| \mid D\in \mathfrak{G}(n,k,\ell)\}$$ and
$$g(n,k,\ell)=\min\{|A(D)| \mid D\in \mathfrak{G}(n,k,\ell)\}.$$
We further define $$Ex'(n,k,\ell)=\{D\mid D\in
\mathfrak{G}(n,k,\ell), |A(D)|=G(n,k,\ell)\}$$ and
$$ex'(n,k,\ell)=\{D\mid D\in \mathfrak{G}(n,k,\ell),
|A(D)|=g(n,k,\ell)\}.$$

Sun and Gutin \cite{Sun-Gutin2} gave the following
characterizations.

\begin{pro}\label{pro11}\cite{Sun-Gutin2}
The following assertions hold:\\
$(i)$~A digraph $D$ is minimally strong subgraph
$(k,1)$-arc-connected
if and only if $D$ is minimally strong digraph;\\
$(ii)$~Let $2\leq k\leq n$. If $k\not\in \{4,6\}$,~or,~$k\in
\{4,6\}$~and~$k<n$, then a digraph $D$ is minimally strong subgraph
$(k,n-1)$-arc-connected if and only if $D\cong
\overleftrightarrow{K}_n$.
\end{pro}

\begin{thm}\label{thm18}\cite{Sun-Gutin2}
A digraph $D$ is minimally strong subgraph $(2,n-2)$-arc-connected
if and only if $D$ is a digraph obtained from the complete digraph
$\overleftrightarrow{K}_n$ by deleting an arc set $M$ such that
$\overleftrightarrow{K}_n[M]$ is a union of vertex-disjoint cycles
which cover all but at most one vertex of
$\overleftrightarrow{K}_n$.
\end{thm}

Sun and Jin characterized the minimally strong subgraph
$(3,n-2)$-arc-connected digraphs.

\begin{thm}\label{thm19}\cite{Sun-Jin}
A digraph $D$ is minimally strong subgraph $(3,n-2)$-arc-connected
if and only if $D$ is a digraph obtained from the complete digraph
$\overleftrightarrow{K}_n$ by deleting an arc set $M$ such that
$\overleftrightarrow{K}_n[M]$ is a union of vertex-disjoint cycles
which cover all but at most one vertex of
$\overleftrightarrow{K}_n$.
\end{thm}

Theorems \ref{thm18} and \ref{thm19} imply that the following
assertions hold: $(i)$~For $k\in \{2,3\}$,
$Ex'(n,k,n-2)=\{\overleftrightarrow{K_n}-M\}$ where $M$ is an arc
set such that $\overleftrightarrow{K}_n[M]$ is a union of
vertex-disjoint cycles which cover all but exactly one vertex of
$\overleftrightarrow{K}_n$. $(ii)$~For $k\in \{2,3\}$,
$ex'(n,k,n-2)=\{\overleftrightarrow{K_n}-M\}$ where $M$ is an arc
set such that $\overleftrightarrow{K}_n[M]$ is a union of
vertex-disjoint cycles which cover all vertices of
$\overleftrightarrow{K}_n$.

Sun and Jin completely determined the precise value for
$g(n,k,\ell)$. Note that $(n,k,\ell)\not\in \{(4,4,3),
(6,6,5)\}$ by Theorem \ref{thm12} and the definition of
$g(n,k,\ell)$.

\begin{thm}\label{thm20}\cite{Sun-Jin} For any triple $(n,k,\ell)$ with $2\leq k\leq n, 1\leq \ell \leq n-1$ such that $(n,k,\ell)\not\in \{(4,4,3), (6,6,5)\}$, we have
$$g(n,k,\ell)= n\ell.$$
\end{thm}

Some results for $G(n,k,\ell)$ were obtained as well.

\begin{pro}\label{pro12}\cite{Sun-Jin}
We have $(i)$~$G(n,n,\ell)\le 2\ell(n-1)$; $(ii)$~For every $k$
$(2\le k\le n)$, $G(n,k,1)=2(n-1)$ and $Ex'(n,k,1)$ consists of
symmetric digraphs whose underlying undirected graphs are trees;
(iii) $G(n, k, n-2)=(n-1)^2$ for $k\in \{2,3\}$.
\end{pro}

Note that the precise values of $g(n,k,\ell)$ for
each pair of $k$ and $\ell$ and the precise values of $G(n,k, n-2)$
for $k\in \{2, 3\}$ were determined. Hence, similar to problems
\ref{op3} and \ref{op4}, the following problems are also
interesting.

\begin{op}\label{op7}\cite{Sun-Jin}
Determine $G(n,k,n-2)$ for every value of $k\ge 2$.
\end{op}

\begin{op}\label{op8}\cite{Sun-Jin}
Find a sharp upper bound for $G(n,k,\ell)$ for all $k\ge 2$ and
$\ell\ge 2$.
\end{op}


\vskip 1cm

\noindent {\bf Acknowledgements.} Yuefang Sun was supported
by National Natural Science Foundation of China (No.11401389) and China Scholarship Council (No.201608330111).
Gregory Gutin was partially supported by Royal Society Wolfson Research Merit Award. 

\end{document}